\documentclass[10pt,twoside,reqno]{amsart}
\usepackage{mathbbol,mathtools,slashed}

\usepackage{multirow}
\usepackage[bookmarksnumbered, plainpages]{hyperref}
\usepackage[rightcaption]{sidecap}
\usepackage{caption,xparse}
\usepackage{pgfplots}

\usepackage{float}

\usepackage[english]{babel}
\usepackage[latin1]{inputenc}
\usetikzlibrary{positioning,arrows,patterns}

\usepackage{graphicx}
\usepackage{epstopdf}
\usepackage{subfig}
\usepackage{color}
\usepackage{amsthm}
\usepackage{amsmath, amsfonts, amssymb}
\usepackage[figuresright]{rotating}
\begin{document}
   \title{Critical Temperature in Nonextensive Hadron Models: A Pressure-based Approach}
   \author{\small M. A. A. AHMED$^{1,\dag}$
    H. ZAINUDDIN$^{2,\ddag}$ }
\thanks{\scriptsize emails: \\ $^{\dag}$ mohammed\_h7@$_{\text{yahoo.com}}^{\text{taiz.edu.ye}}$.\\ $^\ddag$ Hishamuddin.zainuddin@xmu.edu.my.}
\maketitle
\begin{center}
\address{
 $^{1}$ Physics Department, Faculty of Science, Taiz University Al-Turba branch, Taiz, Yemen.\\
 $^{2}$ Department of Physics, Faculty of Science, Universiti Putra Malaysia (UPM), Selangor, Malaysia.}
\end{center}
\begin{abstract}
The determination of the critical temperature $T_c$ in strongly interacting matter is central to understanding the transition between confined hadronic matter and deconfined quark-gluon degrees of freedom. In high energy nuclear collisions, intrinsic fluctuations, memory effects, and long-range color interactions can drive the fireball evolution away from purely Markovian thermalization, making nonextensive statistical descriptions relevant near equilibrium. We investigate the critical temperature of nonextensive hadron models using a pressure-based formulation derived from Tsallis entropy. Here we show that the qualitative structure of the pressure equation is controlled by both the nonextensive parameter $q$ and the model-dependent bag contribution $B^{\text{model}}$. For $q=0$, the pressure equations are monotonic and admit a unique positive root, whereas for $q=1$ the seventh-order term vanishes and analytic solutions can be obtained. For $q>1$, the negative $T^7$ contribution modifies the pressure behavior and can generate multiple positive roots at finite volume. 
For $q\geq1$, $T^{\text{model}}_c$ is $\left(102.3\leq T_c^{\text{MIT}}\leq106.1,\ 86.5\leq T_c^{\text{Dual}}\leq 87.8,\ \text{and} \ 113.1\leq T_c^{\text{Fuzzy}}\leq 121.1\right)$ MeV. At $q=0$, $T^{\text{model}}_c \sim 55$ MeV marks the nuclear liquid-gas transition. A weighted comparison constrained by the MIT bag model at $q=1$ gives $(T_c^{\text{w}}\approx102)\ \text{MeV}$, with Fuzzy and Dual QCD terms entering as model-propagation corrections. The interval $(0.99\leq q\leq 1.002)$ therefore provides the most reliable domain for a common physical interpretation and a composite estimate of the critical temperature.
\end{abstract}
\markboth{\rightline {\sl M.A.A. Ahmed \& H. Zainuddin}}
        {\leftline{\sl $T_c$ in Nonextensive Hadron Models: A Pressure-based Approach}}

\section{\textbf{INTRODUCTION}}\
Quantum chromodynamics (QCD) is the non-Abelian gauge theory of the strong interaction, in which quarks and gluons carry color charges and they exhibit two defining emergent properties, namely, the confinement at long distances and asymptotic freedom at short distances \cite{Gross2023}. This scale dependence implies that, when nuclear matter is heated or compressed enough, hadrons are no longer the optimal degrees of freedom, and QCD predicts a crossover to a deconfined, strongly interacting quark-gluon plasma (QGP) \cite{Gross2023,Gross1973}. The latter can be studied experimentally in Relativistic Heavy Ion Collider (RHIC) and theoretically with lattice QCD, effective field theory, and hydrodynamics \cite{Gross1973,Politzer1974,Fodor2014,Ding2015}. Rather than behaving like a weak gas of free partons, the QGP created at RHIC and the Large Hadron Collider (LHC) is well described as a nearly perfect liquid with rapid equilibration, strong collective flow, and substantial jet-energy loss, and thus making it a precision laboratory for emergent many-body dynamics in QCD under extreme conditions \cite{Busza2018,Schenke2020,Arslandok2023}. Notably, particle yields in central collisions at RHIC were analyzed by comparing Boltzmann-Gibbs (BG) statistics with the Tsallis statistics (TS) \cite{Shao2010,Marques2015,Kyan2022}.

In this context, the parameter $q$ is not extracted from experimental fittings \cite{Vanesa2025}. Instead, it is defined through the nonextensive entropy formulation of the quark-gluon system, directly incorporating its effect on the pressure equation of state \cite{Vanesa2025,Tsallis1988}. Maintaining perturbative stability and root continuity within a finite-volume ideal gas approximation restricts our analysis to $q \approx 1$ \cite{Lima2020}. Because of our entropy definition, chosen observables, and theoretical framework differ fundamentally from fractal approaches, where $q$ is inferred from mass or transverse-momentum spectra. The resultant values of $q$ across the two approaches are not directly comparable \cite{Marques2015,Deppman2012,Deppman2014,Deppman2018,Andrade2020}.

The connection between some quasi-stationary states and $q-\text{Gaussians}$ distributions has been established before. The role played by $q-\text{Gaussians}$ in TS is analogous to that of Gaussians in BG statistics, where BG statistics is a specific instance of $q\longrightarrow1$ \cite{Tsallis1988}. Through TS, one can parametrize deviations from BG statistics often associated with correlations, memory effects, or incomplete equilibration within a formal framework that reduces to ordinary equilibrium thermodynamics at $q$ equals unity \cite{Tirnakli2016,Tsallis2009}. The system recovers the BG limit at $q=1$ and exhibits non-extensive Tsallis behavior at $q>1$. At the Tsallis limit $q=0$, it reduces to the Hartley entropy. In principle, the entropy parameter $q$ is determined by the system dynamics, which is typically too complex to compute from first principles, although in certain cases $q$ can be related to microscopic or mesoscopic quantities. For many-body physics, a relation between the particle number $(N)$ and Tsallis parameter has been given in the form $q-1=\frac{n}{N}$ ($n$ is constant), together with the observation that the 2-particle correlations weaken as $N\rightarrow\infty$, i.e., as the system approaches molecular chaos \cite{Lima2020}. These results help clarify the respective domains of validity of BG and its nonextensive generalization TS \cite{Tirnakli2016}.

In this work, TS uses the $q$-parameter as an extension parameter of BG thermodynamics, with the standard extensive limit recovered at $q=1$. The deviations from unity are interpreted phenomenologically as effective signatures of correlations, finite size effects, fluctuations, or incomplete equilibration. Accordingly, $q$ is treated as a system-dependent and observable parameter rather than a universal QCD constant \cite{Lima2020,Astorga2024,Vanesa2025}. Our treatment is phenomenological, since $q$ is introduced through the Tsallis entropy and pressure equation rather than derived directly from QCD. This differs from self-consistent and fractal approaches, where nonextensivity is related to self-similarity, hierarchical structure, scale invariance, or the hadronic spectrum \cite{Deppman2012,Deppman2014,Deppman2018,Andrade2020}. Therefore, the numerical values of $q$ need not coincide, since they are constrained by different physical inputs and observables. The two approaches are complementary: microscopic models address the possible origin of nonextensivity, while the present work studies its effect on the pressure and transition temperature.

In QCD, confinement refers to the low-temperature, low-density regime in which quarks and gluons are not observed as isolated particles but are bound into color-singlet hadrons \cite{Gross1973,Huang2023}. In contrast, deconfinement occurs when temperature or baryon density becomes large enough that the relevant thermodynamic degrees of freedom change to quarks and gluons and the system crosses into quark matter or a QGP \cite{Lugones2023}. In thermodynamic theory, this change is tracked through the Equation of State (EoS), especially the pressure, which is suppressed in the confined phase and rises rapidly once deconfined degrees of freedom are activated \cite{Fraga2024}. While in bag-type descriptions the bag constant $B$ represents the vacuum-energy difference between the nonperturbative confined vacuum and the perturbative deconfined phase, so that it enters as a negative shift in the system pressure, $P_{\text{Total}}\sim P_{\text{Partons}}+P_{\text{Vac}}$, delaying deconfinement until the matter pressure overcomes this vacuum pressure and making $P_{\text{Vac}}$ a phenomenological measure of confinement effects in effective QCD models \cite{Chodos1974,DeGrand1975}.

In these regimes, phenomenological models play an important complementary role. Among them, the MIT bag model offers a simple yet physically transparent description of quark confinement by introducing a vacuum pressure, the bag constant $B_{\text{MIT}}$, that balances the internal thermal pressure of the partonic \cite{DeGrand1975,Philipsen2013}. Despite its simplicity, this model has been widely used to study the partonic EoS, deconfinement transition, and partonic matter properties at finite temperatures \cite{Karsch2005,Cleymans1986,Gasser1987}. However, most applications of the MIT bag model assume a thermodynamic limit, while systematic studies of finite size corrections remain relatively limited \cite{Teweldeberhan2003}. Besides that, some studies have shown discrepancies when comparing them with the bag model predictions. The surprise of these analyses is that at zero chemical potential, it reveals a term quadratic in temperature as the leading correction to the ideal gas term in the pressure \cite{Andreev2007}. This suggests a Fuzzy bag model $B_{\text{Fuzzy}}$ for the analogous pressure in QCD, with dynamical quarks, which is equivalent to the pressure proposed by Pisarski \cite{Pisarski2007}. On the other hand, the QGP transition is modeled using a bag constant defined by the radial pressure of vector glueball masses in a magnetically condensed vacuum that is introduced within the dual QCD framework of magnetic symmetry \cite{Chandola2009,Punetha2022}. By evaluating the partition function, this approach identifies critical temperatures where the phase dynamics shift: a fixed bag constant $B_{\text{Dual}}$ indicates a first-order transition, whereas a thermal bag constant $B_{\text{Dual}}(T)$ triggers a rapid crossover \cite{Punetha2022,Chandola2016}.

Nonextensive extensions of the bag model based on TS has successfully reproduced standard hadron masses and has naturally been extended to exotic multiquark states \cite{Astorga2024,Vanesa2025,Dragoa2004,Lavagno2010}. In addition, incorporating a chemical potential dependent bag constant can significantly changes neutron star matter, allowing an isentropic hadronic and partonic phase transition. This modification ultimately yields stable partonic cores and the existence of massive compact stars compatible with current observations \cite{Joshi2021,Hasmik2025}. Henceforth, this current study presents the pressure of the QGP system within TS for $q\geq1$, by considering vacuum pressure for three phenomenological models: MIT, Fuzzy, and Dual.

The MIT, Fuzzy, and Dual models are not treated as simultaneously coexisting phases but as alternative effective descriptions of the nonperturbative contribution to the QCD pressure. Their common structure can be written schematically as
$$P_q^{\text{Model}}(T,V,\mu)=P_q^{\text{Matter}}(T,V,\mu)+P^{\text{(Model Bag)}}(T,\mu).$$
In the MIT bag model, the vacuum contribution is represented by a constant bag pressure, whereas the Fuzzy bag model introduces a lattice-motivated nonperturbative $T^2$ correction to the leading $T^4$ behavior. Meanwhile, the Dual-QCD model relates the vacuum pressure to magnetic symmetry breaking and vector-glueball modes within the magnetically condensed QCD vacuum \cite{DeGrand1975,Andreev2007,Pisarski2007,Chandola2016}. The three models can therefore be compared using the same thermodynamic pressure criterion, while the differences among their predicted $T_c$ values quantify the model dependence associated with the treatment of the nonperturbative QCD vacuum.
\section{Tsallis Entropy and Pressure}\label{sec:1}
In a grand canonical ensemble, we can get the expressions of entropy coming from each component \cite{Swendsen2017a}. We can then calculate the pressure using the Maxwell relations of thermodynamics by deriving $S_T$ with respect to $V$ and then integrating the expression with respect to temperature $T$ \cite{Swendsen2017b}. By incorporating Tsallis entropy in the phenomenological bag model, we lose the independent contribution of the separate subsystems to the total pressure \cite{Astorga2024,Vanesa2025}.  We can no longer isolate the pressure induced by quarks from the pressure induced by gluons \cite{Astorga2024}. This makes sense, given that protons are a complex and intertwined mixture of interacting quarks and gluons. Other methods can separate these components \cite{Vanesa2025}.
\subsection{Tsallis Entropy}\label{sec:11}
Tsallis proposed a generalization of BG statistics in his work \cite{Tsallis1988,Tsallis2009}. While entropy in classical statistical mechanics is considered an extensive property, Tsallis's statistics take into account systems that may not satisfy such extensivity condition \cite{Swendsen2017a,Swendsen2017b}. Instead, non-unversal entropy is defined as follows:
\begin{equation}\label{eq}
S_T = k_B \frac{1 - \sum_{i=1}^{\omega} \varrho_i^q}{q - 1},
\end{equation}
where $q$ is a real number characterizing a non-extensive system, $\omega$ is the number of microscopic states in the system, and $\varrho_i$ is the probability of finding states in the system. In eq. (\ref{eq}), $k_B$ is a positive constant, which is, in fact, Boltzmann's constant $k_B$ as in classical statistical mechanics. For three independent systems, A, B, and C, their combination can be represented by the following formula:
\begin{equation}\label{eq0}
\begin{aligned}
S_T(A,B,C)={}&S_A + S_B + S_C
+ (1-q)\sum_{i<j} \alpha_{ij} S_i S_j \\
&+ (1-q)^2 \alpha_{ABC} S_A S_B S_C .
\end{aligned}
\end{equation}
We note that the triple system leads to the Tsallis-type entropy in this system up to the second order of the q-parameter. Now, let us assume that ($A$, $B$, and $C$) represent the quark, gluon, and vacuum systems, respectively.
\begin{equation}\label{eq1}
\alpha_{AB}=1,\quad \alpha_{AC}=\alpha_{BC}=0, \quad \alpha_{ABC}=0.
\end{equation}
In the present model, only the quark-gluon sector is allowed to be non-extensive, while the vacuum is treated as an independent background. Accordingly, ($A$ and $B$) correlate, whereas ($A$ and $C$) are independent, and ($B$ and $C$) are also independent. In other words, no genuine three-system correlation among ($A$, $B$, and $C$) is included. These assumptions are implemented through the selection rules. Hence, the generalized Tsallis entropy reduces to
\begin{equation}\label{eq2}
S_T(A,B,C)=S_A+S_B+S_C+(1-q)\,S_A S_B.
\end{equation}
The entropy of the vacuum is neglected because the vacuum contributes nothing to entropy, i.e., $S_C=0$, the above expression further simplifies to
\begin{equation}\label{eq3}
S_T=s_q+s_g+(1-q)\,s_q\ s_g.
\end{equation}
The system of ultra-relativistic fermion gas, such as the quark entropy \cite{KAPUSTA}, is given as follows:
\begin{equation}\label{eq4q}
    s_{q} = g_q \left[ \frac{7\pi^2}{90} + \frac{1}{6} \left( \frac{\mu}{T} \right)^2 \right] VT^3.
\end{equation}
The system of ultra-relativistic boson gas, such as the entropy of gluons \cite{KAPUSTA}, is given as follows:
\begin{equation}\label{eq4g}
    s_{g} = \frac{4g_g \pi^2}{90} VT^3.
\end{equation}
We get the Tsallis entropy of the quark-gluon system when substituting Eqs.(\ref{eq4q}) and (\ref{eq4g}) into Eq.(\ref{eq3}) \cite{Vanesa2025}, resulting in the explicit form for the Tsallis entropy $S_T(T,\mu;V,q)$ as:
\begin{equation}\label{eq6}
\begin{aligned}
S_T(T,\mu;V,q)={}&
\left[
\frac{74\pi^2}{45}
+2\left(\frac{\mu}{T}\right)^2
\right]VT^3 \\
&+
\frac{128\pi^2}{15}(1-q)
\left[
\frac{7\pi^2}{90}
+\frac{1}{6}\left(\frac{\mu}{T}\right)^2
\right]
(VT^3)^2 .
\end{aligned}
\end{equation}
In this formula, $T$ and $\mu$ represent thermal variables, $V$ is a mechanical variable, and $q$ is a nonextensive statistical parameter describing deviations from standard extensive thermodynamic behavior. This study considers up and down quarks as two flavors, possessing three color charges and two spin orientations; and hence multiplicity factor in eq.(\ref{eq4q}) is expressed as ($g_q=2 \times 3 \times 2$). Similarly for gluons, the multiplicity factor in eq.(\ref{eq4g}) is ($g_g=2\times 2^3$), where the factor of eight is due to the generators of the $SU(3)$ color group and the factor of two is due to spin orientation.\\
We then determine the thermal pressure using Maxwell relations by differentiating $S_T$ with respect to volume and integrating the resulting expression over temperature.
\begin{equation}\label{eq7}
\left( \frac{\partial S_T}{\partial V} \right)_{T,\mu}=\left( \frac{\partial P_T}{\partial T} \right)_{V,\mu},
\end{equation}
we integrate over temperature to obtain
\begin{equation}\label{eq8}
P_T(T,\mu) =
\int \left( \frac{\partial S_T}{\partial V} \right)_{T,\mu} dT
+ \mathcal{C}(V,\mu).
\end{equation}
The integration constant is determined by matching to the extensive limit. Equation (\ref{eq7}) shows the relationship between the derivatives of entropy and pressure. The pressure $P_T$ used in this work is the EOS pressure at equilibrium. For a homogeneous system, the corresponding entropy density satisfies the relation $S_T = (\partial P_T/\partial T)_\mu$, and this derivative is calculated at a constant chemical potential \cite{KAPUSTA}.
\subsection{Tsallis Pressure}\label{sec:12}
We shall consider only bags with static conditions for (quark, antiquark, and gluon) systems with no interaction, with an equal number of quarks and antiquarks. At high energies, the gluons behave like an ultra-relativistic Bose gas. Therefore, the pressure emerging from the gluons is given by:
\begin{equation}\label{eq9}
P_g (T)= g_g \frac{\pi^2}{90} T^4,
\end{equation}
We also take into account for the continuous creation and annihilation of quark-antiquark pairs by treating the system as a mixed gas. Even though these processes are ongoing, we focus on the net difference between quarks and antiquarks rather than the specific remnants of individual reactions. This allows us to establish a direct relation between the particle count and their respective chemical potentials. Consequently, the quark pressure of the system is expressed as:
\begin{equation}\label{eq10}
P_q (T,\mu)=
\frac{g_q}{24 \pi^2} \mu^4
+ \frac{g_q}{12} \mu^2 T^2
+ \frac{7 \pi^2}{720} g_q T^4.
\end{equation}
The total pressure of the system is:
\begin{equation}\label{eq12}
P_T(T,\mu;V,q)
= P_g(T) + P_q(T,\mu) + P_{\text{Tsallis}}(T,\mu;V,q)+ C(V,\mu).
\end{equation}
The total pressure contains the term standard ideal gas $T^4$, terms of gluons and quarks, the $\left(\frac{\mu}{T}\right)^2$ and $\mu^4$ quarks contributions, and a nonextensive correction $(1-q)V T^7$ encoding quark-gluon correlations in this model. Here $C(V, \mu)$ is an integration constant, but note it has no $T$ dependence after the integration because it is the ``constant'' of integration with respect to $T$ \cite{Vanesa2025}. The total pressure is written in eq.(\ref{eq12}), where $P_{\text{BG}}(T,\mu)=P_g(T) + P_q(T,\mu)$ represents the BG pressure, and $P_{\text{Tsallis}}(T,\mu;V,q)$ is the Tsallis pressure contribution, which is given by
\begin{equation}
\frac{P_{\text{Tsallis}}(T,\mu;V,q)}{T^4}=  \frac{2\pi^4}{2025} g_q g_g (1-q) \left[1 + \frac{3}{\pi^2} \left( \frac{\mu}{T} \right)^2 \right] V T^3.
\end{equation}

\begin{figure}[h]
 \hspace{-0.75cm}
 \includegraphics[width=450pt]{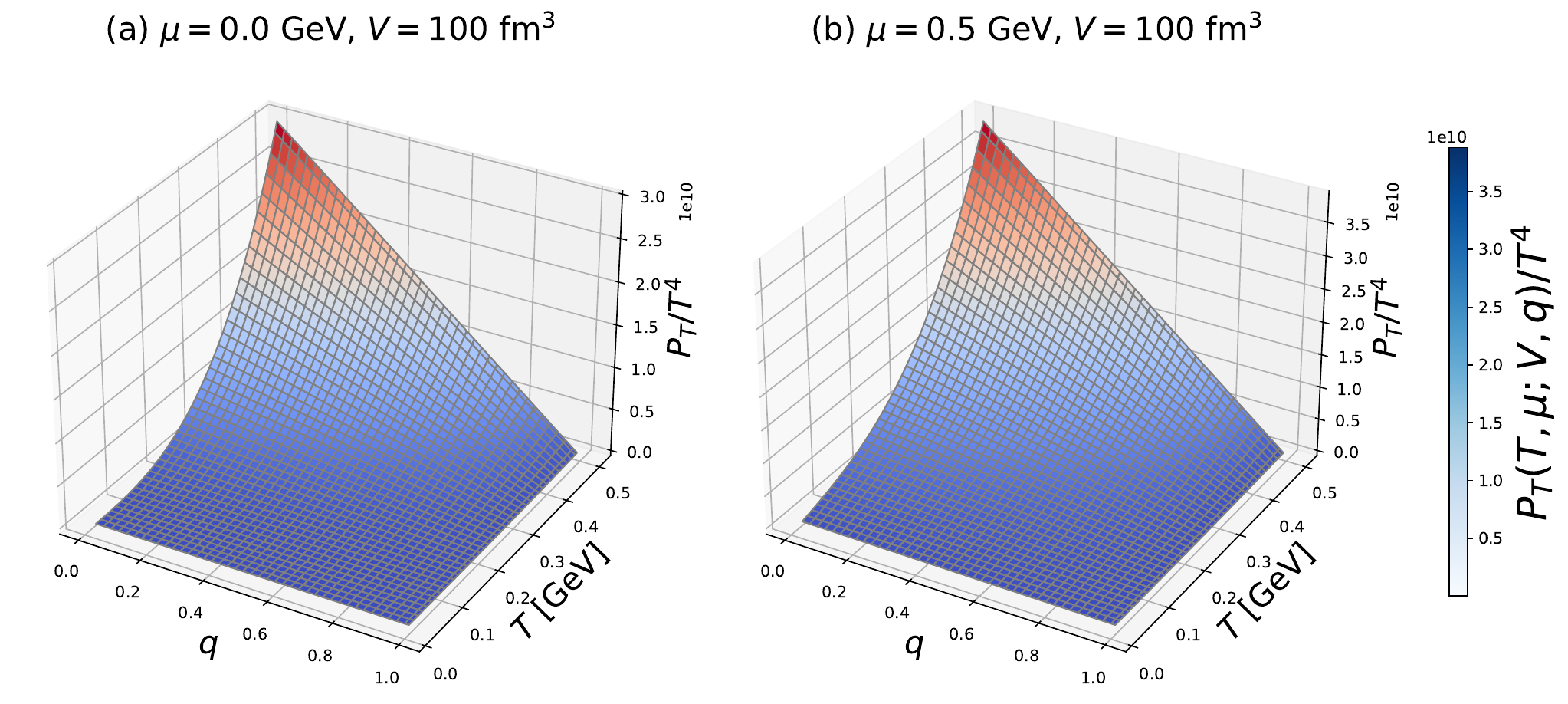}
  \caption{The plots of the pressure $P(T,\mu;V,q)/T^{4}$ versus $q$ and $T$ for $V=100\ \mathrm{fm}^{3}$: (a) $\mu=0$ and (b) $\mu=0.5\ \mathrm{GeV}$. The comparison shows the effect of finite chemical potential on the thermodynamic behavior of the system within the nonextensive framework}\label{figb1}
\end{figure}
The $V T^3$ scaling is a hallmark of non-extensivity. In standard BG statistics ($q=1$), entropy must be extensive ($S_{\text{BG}} \propto V T^3$). The appearance of $(S_{\text{Tsallis}}\propto (VT^3)^2)$ indicates that the system's degrees of freedom are correlated across the entire volume of the ``bag,'' a feature that the Tsallis model uses to explain the mass spectra of complex hadrons like tetraquarks \cite{Vanesa2025}. The integration constant is given by
\begin{equation}
\mathcal{C}(V,\mu)= \frac{g_q}{24 \pi^2} \mu^4.
\end{equation}
So, in this model, $\mathcal{C}$ is independent of ($V$ and $T$) and ensures the correct $\mu^4$ piece appears when $q=1$.
At $q=1$, the non-extensive correction term vanishes. The pressure $P_T(T,\mu=0;V,q=1)$ represents the standard Stefan-Boltzmann (SB) limit for a massless gas at zero chemical potential \cite{Mhamed2021}, while $P_T(T,\mu;V,q=1)$ includes the quadratic correction for finite baryon density $(\mu/T)^2$. The difference between them, $\Delta P_1=\left(\frac{P_T(T,\mu;V,q=1)-P_T(T,\mu=0;V,q=1)}{T^4}\right)$, separates the pure contribution of the chemical potential to the normalized pressure. Non-extensive framework $\Delta P_0=\left(\frac{P_T(T,\mu;V,q=0)-P_T(T,\mu=0;V,q=0)}{T^4}\right)$: At $q=0$ (the maximum deviation in this range), the system incorporates long-range correlations and temperature fluctuations \cite{Mhamed2026,Mhamed2026a}. $P_T(T,\mu=0;V,q=0)$ shows that even at $\mu=0$, the pressure is significantly boosted by a factor proportional to $VT^3$. The effect of amplification is $\Delta P_0$ versus $\Delta P_1$, and the most significant result is the difference in $\Delta P_0$. In the standard case, the density contribution depends only on the ratio $\mu/T$. However, in the non-extensive case, the effect of the chemical potential is multiplied by the volume of the system and temperature. This physically suggests that in non-extensive QGP, the presence of baryon density generates a much higher pressure than predicted by standard models, due to the correlation between global fluctuations $q$ and particle density. Now we can enter the pressure ratio $\mathcal{P}_q(T;V)$, which will be included as a dimensionless quantity
\begin{equation}
\mathcal{P}_q(T;V)=\frac{P_{\text{Tsallis}}(T;V,q)}{P_{\text{BG}}(T)}
=\frac{4\pi^2}{1665}g_q g_g(1-q)\ VT^3.
\end{equation}
Then the total pressure becomes
\begin{equation}\label{pt:pg}
  \frac{P_T(T;V,q)}{P_{\text{BG}}(T)}=\begin{cases}
                                        \displaystyle 1+\mathcal{P}_q(T;V), &  \mbox{if }\ \ \  \mathcal{P}_q(T;V)\gg 1, \\
                                        \\
                                        \displaystyle \frac{1}{1-\left|\mathcal{P}_q(T;V)\right|}, &  \mbox{when }\ \ \ \left| \mathcal{P}_q(T;V)\right| \ll 1.\\
                                      \end{cases}
\end{equation}
The perturbative or small-correction regime is known when $\left|\mathcal{P}_q(T;V)\right|\ll 1$. Therefore, the correction is controlled by the single combination $(1-q)VT^3$. This determines the geometry of the perturbative domain in parameter space: the approximation is reliable only when this dimensionless quantity remains small. The resumed form has a pole when
\begin{equation}
1-\mathcal{P}_q(T;V)=0.
\end{equation}
This works out to give the pole temperature:
\begin{equation}
T_{\text{pole}}^3 = \frac{1665}{4\pi^2 g_g g_q}\frac{1}{(1-q)V}.
\end{equation}
The numerical evaluation of the pole temperature, $T_{\text{pole}}$, within the sub-extensive range $0.01 < q < 0.99$ at a fixed volume $V = 100 \text{fm}^3$, shows that the critical values fall within the interval:
\begin{equation}\label{pole}
0.0123 ~\text{GeV} < T_{\text{pole}} < 0.119 \ \text{GeV}.
\end{equation}
In the range $q<1$, the appearance of this pole at low to intermediate temperatures, as shown in eq.(\ref{pole}), indicates a fundamental limit for the stability of the thermal gas. In contrast, for $q>1$, the analytical framework does not predict any physical variation for $T_{\text{pole}}$. This suggests that the super-extended state is inherently more stable than the pressure-driven singularities that characterize the sub-extensive regime in this model. However, it is not a singularity point for the original summation pressure $P_T(T;V,q)/T^4$, which remains bounded for all values of $q$. The pole appears only in the resumed representation and should, therefore, be interpreted as a signal of the breakdown of the small-correction expansion. The regime $\left|\mathcal{P}_q(T;V)\right|\ll 1$ defines the domain where the correction is perturbative.
\begin{equation}
\frac{P^{\text{res}}_T(T;V,q)}{T^4}=\frac{\frac{37}{90}\pi^2}{1-\frac{256}{555}\pi^2(1-q)\ VT^3}.
\end{equation}
The quantity $P^{\text{res}}_T(T;V,q)$ should be understood as a geometric resummation introduced solely as a formal indicator of the perturbative domain, rather than as a physical pressure. The pole at $\lvert\mathcal{P}_q(T;V)\rvert=1$ signifies the loss of convergence of the auxiliary geometric series and the limit of the perturbative description, rather than a genuine divergence, singularity, or phase transition of the original EOS.
\begin{figure}[h]
 \hspace{-0.75cm}
 \includegraphics[width=450pt]{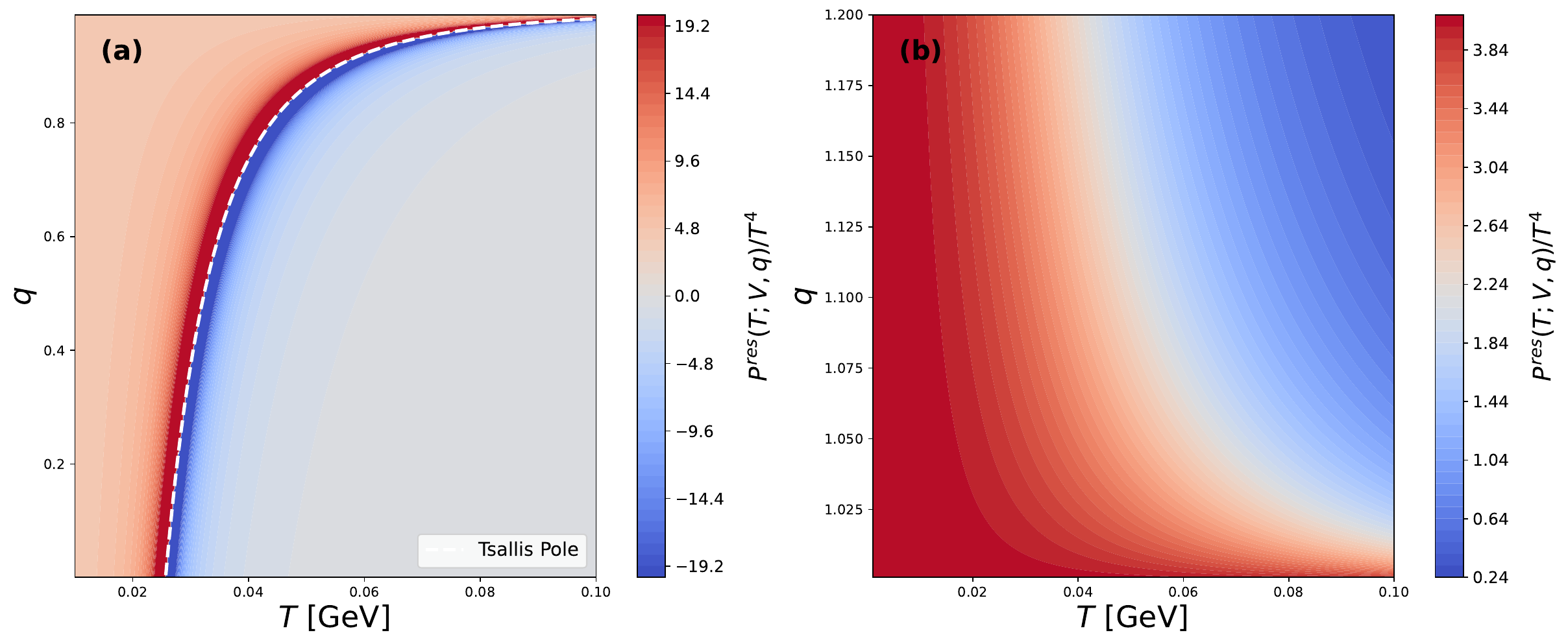}
  \caption{Pressure plots $P^{res}(T;V,q)/T^{4}$ versus $q$ and $T$ at $V=100 \mathrm{fm}^{3}$: (a) The subplot shows a pole in $P^{res}/T^4$, indicated by the white curve, where the denominator of the analytical expression vanishes. The contour lines converge near this boundary, indicating a marked increase and change in the signal across the single singular. (b) The entire subplot lies within a regular range, with no poles present within this range. The pressure remains confined and decreases smoothly with temperature.}\label{figb11}
\end{figure}
The dimensionless ratio in eq.(\ref{pt:pg}) characterizes the interplay between the extensive BG background and the non-extensive Tsallis correction. In the perturbative regime ($\mathcal{P}_q(T;V) \ll 1$), the system occupies a quasi-equilibrium state dominated by BG $T^4$ scaling, where the correction term acts marginally and can be approximated as $e^{\mathcal{P}_q(T;V)}$. Conversely, the asymptotic limit ($\mathcal{P}_q(T;V) \gg 1$) marks a profound scaling transition. The classical background becomes negligible as the $T^7$ non-extensive power law supersedes standard SB-like physics. This duality underscores a shift from a locally intensive thermal gas to a globally correlated system where collective dynamics and long-range correlations govern the EOS.
\section{Phenomenological Hadron Models}\label{sec:2}
As is well known, perturbative QCD provides a reliable description at asymptotically high densities and temperatures, where the QCD coupling becomes small, a property known as asymptotic freedom \cite{Gross2023,Gross1973}. At lower densities and temperatures, one often resorts to the bag model, a phenomenological framework that accounts for nonperturbative strong-interaction effects through a bag constant \cite{Gross2023,Politzer1974,Mameda2023}. The MIT bag model confines quarks in a cavity via a constant vacuum pressure \cite{DeGrand1975,Philipsen2013}, while the Fuzzy bag model restores chiral symmetry by smoothing the bag boundaries \cite{Andreev2007,Pisarski2007}. In parallel, the Dual bag model employs the dual superconductor hypothesis, where magnetic monopolar condensation creates flux pipes to enforce confinement \cite{Chandola2016,Punetha2022}. Together, these models provide phenomenological virtual bridges between perturbative asymptotic freedom and the non-perturbative QCD vacuum.

In the thermodynamic formulation used here, ``Fuzzy'' refers specifically to the non-ideal pressure ansatz motivated by lattice results:
\begin{equation}\label{Pqcd}
P_{\text{QCD}}(T) \simeq f_{\text{pert}}T^4 - B_{\text{Fuzzy}}T^2 - B_{\text{MIT}} + \cdots ,
\end{equation}
which represents a hybrid phenomenological EoS for the QGP at high temperatures. Here, the first term captures the perturbative contribution, the second term accounts for the Fuzzy bag correction, and the final term corresponds to the standard MIT bag constant. This differs from the literal geometric smoothing of the MIT bag boundary.

The wealth of lattice QCD data accumulated in recent years has exposed discrepancies with the predictions of the bag model. In particular, at zero chemical potential, these studies indicate that the leading correction to the ideal-gas contribution to the pressure is quadratic in temperature. This suggests that EOS must be extended by including an additional quadratic term:
\begin{equation}\label{pre}
  P(T;V)\approx a\ T^4+b\ T^2-B\ T^0.
\end{equation}
 In this work, ``critical temperature'' refers to the phenomenological pressure-balance transition temperature, derived from the finite-volume total pressure, obtained when eq. (\ref{pre}) vanishes.
The terms $a\ T^4$ and $b\ T^2$ represent the external thermal pressure of the quarks and gluons, while the Bag term ($B$) provides the internal confinement pressure. The parameters $a$ and $b$ are largely determined by the degrees of freedom of the particles, and the interpretation of $B$ varies considerably depending on the phenomenological model used.
\begin{figure*}[t]
 \centering
 \includegraphics[width=1.0\textwidth]{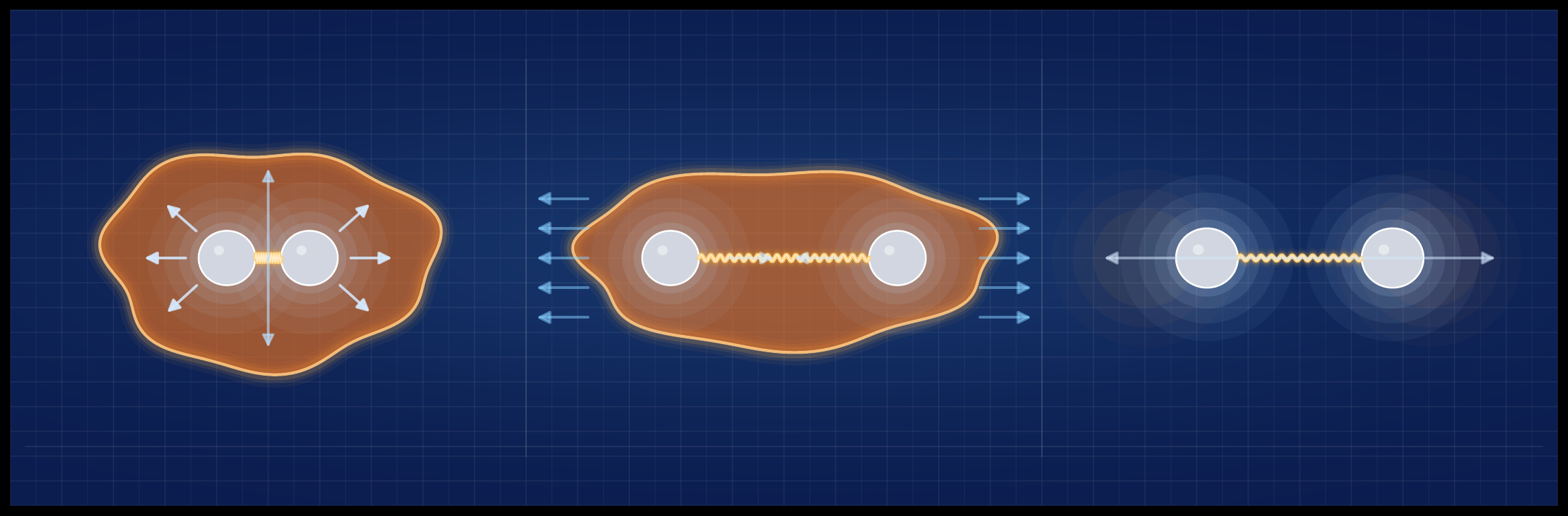}
  \caption{Schematic representation of quarks within the framework of the bag model. The first case shows the quark-confined phase in a compressed bag. And also, the second manifests an expanding confinement region with increasing distance between quarks. While the last appears to be a deconfined phase with a remaining flow connection.}\label{figb2}
\end{figure*}
\subsection{The MIT bag model}\label{sec:21}
In the MIT bag model, confinement is implemented by a constant external ``bag pressure'' $B$. The total pressure of deconfined quarks and gluons is the ideal gas pressure minus $B$.
\begin{equation}\label{MIT}
P_{\text{MIT}}(T;V,q)= P_{\text{GB}}(T)+ P_{\text{Tsallis}}(T;V,q)+P_{\text{(MIT Bag)}}.
\end{equation}
We assumed massless quarks and  gluons, the Tsallis mixed term is exactly as in the normalization. Therefore, in the MIT version of the pressure derived from Tsallis, the vacuum pressure is,
\begin{equation}\label{MIT0}
P_{\text{(MIT Bag)}}=  - B.
\end{equation}
In the eq.(\ref{eq2}), the vacuum entropy is neglected, but the situation is different with pressure, where the vacuum pressure appears as a constant. It is useful here to know the critical temperature, so we assume the following:
\begin{equation}\label{MITc}
    P_{\text{MIT}}(T_c; V,q) = 0.
\end{equation}
The condition given in eq.(\ref{MITc}) represents the pressure equilibrium criterion in the bag model, where the pressure of the confined phase is used as a reference level. More generally, phase equilibrium requires the appropriate thermodynamic potentials to be equal, or equivalently:
\begin{equation}\label{MITc1}
P_{\text{conf}}(T_c) = P_{\text{deconf}}(T_c),
\end{equation}
for homogeneous phases at identical values of $T$. In the statistical description, the probabilities of two confined and deconfined phases being equal, at the thermodynamic limit, correspond to their thermodynamic potential equality, since subextensive finite volume coefficients become negligible. Therefore, the resulting $T_c^{\text{Model}}$ should be interpreted as the effective temperature for deconfinement within the bag model framework rather than as a fundamental temperature of the QCD ordering coefficient, because the current model does not explicitly include the Polyakov loop, center symmetry, or chiral condensate \cite{Philipsen2013,Mameda2023}. The pressure in eq.(\ref{MIT}) contains three conceptually distinct ingredients. The first is the thermal pressure $P_T$, obtained from eq.(\ref{eq7}).
The BG and Tsallis terms appearing in $P_T$ are contributions generated by the same entropy ansatz. The second ingredient is the phenomenological vacuum contribution $P_{\text{bag}}^{\text{Model}}$, specified independently by the MIT, Fuzzy, or Dual model.

It is noted that the MIT model does not depend on temperature, whereas the other two models do. Since this contribution is not derived here from Tsallis entropy, the quantity $P^{\text{Model}}$ is treated as an effective phenomenological pressure rather than a unique microscopic EOS.

\begin{figure}[h]
 \hspace{-0.75cm}
 \includegraphics[width=450pt]{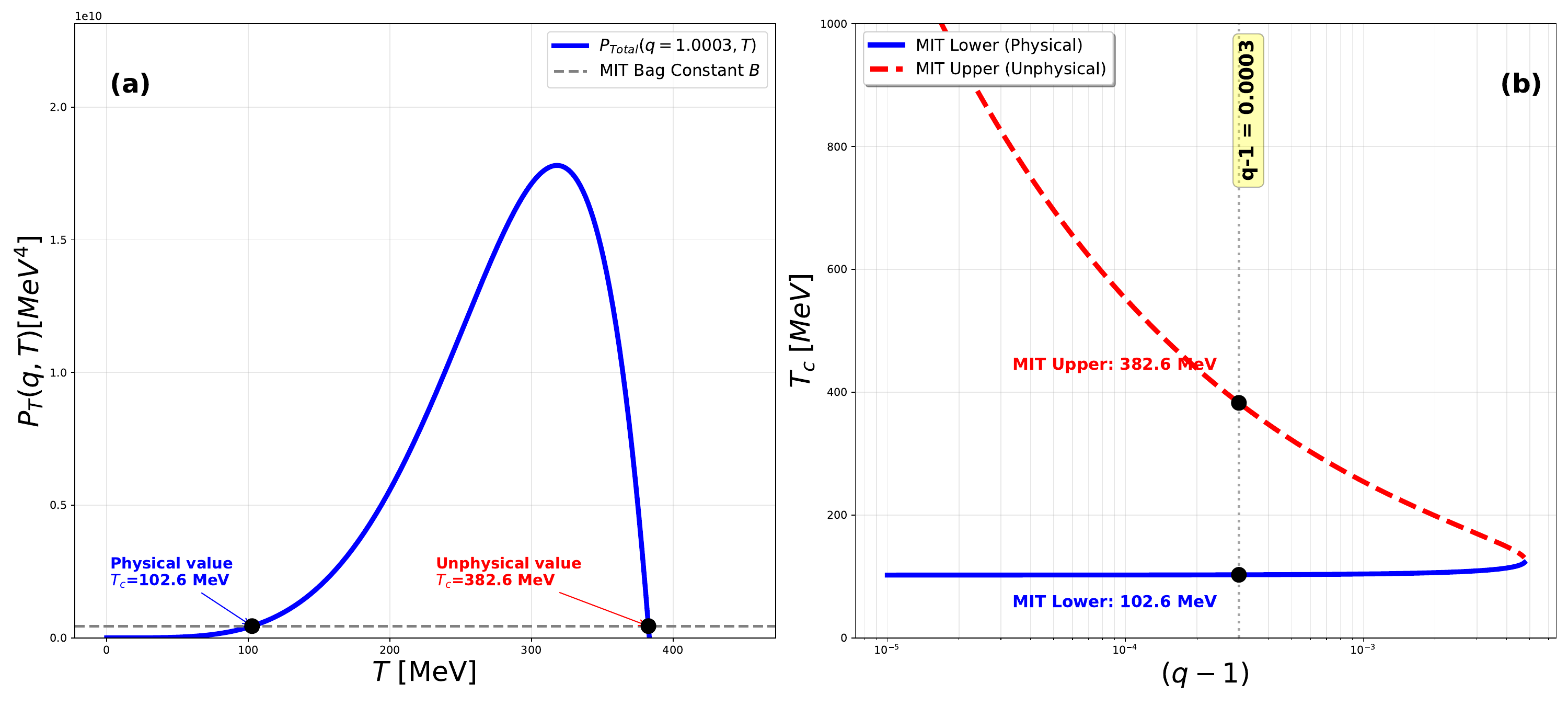}
  \caption{Pressure behavior within the MIT model at $(V=100\ \text{ fm}^3)$. (a) The total pressure $P_T(T,q=1.0003)$, showing branches of solution into physical and unphysical $T_c^{\text{Model}}$. (b) Positive real solutions of $T_c^{\text{Model}}$ as a function of the non-extensivity deviation $(q-1)$, illustrating the dual-branch structure and the approach to the BG limit.}\label{figb4}
\end{figure}

The third ingredient is the prescription used to extract a characteristic temperature. In this work, $T_c^{\text{Model}}$ is defined as a stable positive solution of $P^{{\text{Model}}}(T_c;V,q)=0.$
The pressure balance condition depends on the model and assumes the existence of a zero-pressure reference phase. It includes an explicit term proportional to $(1-q)V$. Thus, it represents a mechanical pressure associated with the finite volume.
Although the general pressure expression remains dependent on $\mu$, the analysis of transitional pressure balance presented below is restricted to $\mu=0$. Accordingly, no general conclusions are drawn here regarding the magnitude of finite $\mu$ effects, as a systematic analysis of the finite density case has been deferred to future work. At $\mu=0$, the ideal part becomes:
\begin{equation}\label{MIT1}
    P_{\text{BG}}(T) = a\ T^4, \quad a = \frac{37\pi^2}{90},
\end{equation}
and the Tsallis pressure is written as follows
\begin{equation}\label{MIT2}
     P_{\text{Tsallis}}(T;V,q) = b (1 - q)  V T^7, \quad b =3\left( \frac{2^4\ \pi}{3\ \sqrt{5}} \right)^4.
\end{equation}
\begin{figure}[h]
 \centering
 \includegraphics[width=350pt]{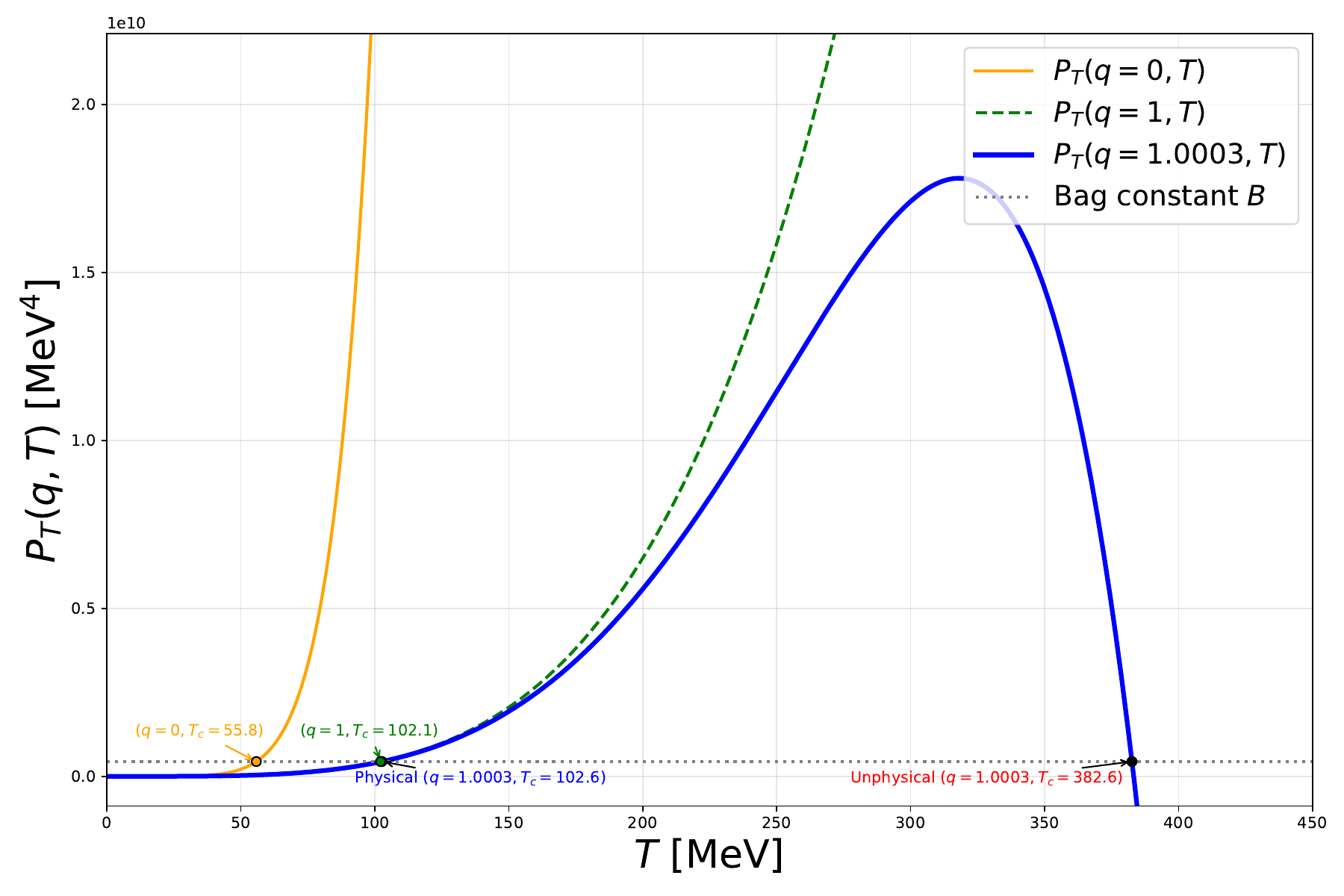}
  \caption{Total pressure $P_T(T,\mu=0)$ as a function of temperature for the MIT Bag Model at $V = 100 \text{ fm}^3$. The plot compares the standard BG statistics $q=1$ with non-extensive cases ($q=0$ and $q=1.0003$). The intersections with $B$ represent $T_c^{\text{MIT}}$; note that for $q > 1$, the negative term of $T^7$ causes the curve to ``bend,'' creating both a physical and an unphysical root.}\label{figb3}
\end{figure}
Within the non-extensive framework QCD bag models, the introduction of the non-exclusive parameter $q > 1$ modifies the thermal pressure through a dominant interaction term proportional to $(1-q)VT_c^7$. Numerically, this modification leads to a branches of solutions space, yielding two distinct positive real values for the critical temperature $T_c$ (see fig.(\ref{figb3})). The lower root in the fig.(\ref{figb4}) represents the physically possible transition point and thermodynamic stability.
On the other hand, the upper root corresponds to an unphysical system where the reaction term grows rapidly, causing the pressure to decrease in a non-uniform manner as the temperature rises. This means that the state is unstable and has a negative entropy density. These two branches converge at a critical stability limit for the $q$-parameter, where the maximum achievable total pressure exactly matches the vacuum bag constant, representing the boundary beyond which the QGP phase can no longer be sustained against the vacuum pressure.\\
By inserting eqs.(\ref{MIT1}) and (\ref{MIT2}) into the expression for total pressure in eq.(\ref{MIT}), and imposing the condition in eq.(\ref{MITc}), we obtain the explicit expression for the constant bag term:
\begin{equation}\label{MIT3}
 a T_c^4 + b (1 - q) V T_c^7-B=0.
\end{equation}
Although lattice-inspired extensions of the bag model are commonly expressed through a quadratic correction, in the present Tsallis-MIT framework the non-ideal contribution is instead encoded in eq.(\ref{MIT3}) via the mixed term $(b(1-q)VT^7)$. This term modifies the pressure and, upon imposing the condition at the critical temperature $T = T_c^{\text{Model}}$, yields the bag constraint.\\
\subsection{The Fuzzy bag model}\label{sec:22}
Now we consider Tsallis and Fuzzy-bag description of QGP thermodynamics, using the same logic via Tsallis pressure and an integration constant, together with the Fuzzy bag pressure ansatz of Pisarski and its thermodynamic integration in Andreev \cite{Vanesa2025,Fraga2009}.
\begin{equation}\label{Fuzzy}
P_{\text{Fuzzy}}(T;V,q)\approx P_{GB}(T)+ P_{\text{Tsallis}}(T;V)+P_{\text{(Fuzzy Bag)}}.
\end{equation}
The Fuzzy Bag pressure is given by
\begin{equation}
P_{\text{(Fuzzy Bag)}}= -( B_{\mathrm{Fuzzy}} T^2+B).
\end{equation}
The usual Fuzzy bag condition applies when $\mu=0$ (this is the standard way to use it with lattice data near $T_c^{\text{Fuzzy}}$). Therefore, we assume the following:
\begin{equation}\label{Fuzzy3}
    P_{\text{Fuzzy}}(T_c;V,q) = 0.
\end{equation}
By expressing the total pressure at $(T=T_c^{\text{Fuzzy}})$, and applying the condition given in the eq.(\ref{Fuzzy3}), we obtain the explicit expression for the constant bag term:
\begin{equation}\label{Fuzzytc}
    a T_c^4 + b (1 - q) V T_c^7 - B^{\frac{1}{2}}_{\text{Fuzzy}} T_c^2-B = 0.
\end{equation}
Equation (\ref{Fuzzytc}) gives the transition pressure-balance condition for the Fuzzy bag model. Here, $B_{\mathrm{fuzzy}}^{1/2}$ is the phenomenological coefficient of the quadratic $T^2$ correction, while $B$ is the temperature-independent vacuum contribution. The $T^2$ term effectively represents nonperturbative deviations from the ideal $T^4$ pressure observed in inspired lattice analyses \cite{Andreev2007,Pisarski2007}.

The quadratic term should be understood as an effective nonperturbative correction to the deconfined pressure, not as evidence for a literal finite-thickness bag surface. In the original Fuzzy bag discussion, the ansatz is intended for a restricted deconfined temperature interval above the vicinity of $T_c^{\text{Fuzzy}}$, before ordinary perturbative behavior becomes dominant \cite{Pisarski2007,Andreev2007}. We use the same pressure structure phenomenologically to compare its critical temperature solution with the MIT and Dual prescriptions.
\subsection{The Dual bag model}\label{sec:23}
We now present the Tsallis pressure formula, replacing the usual phenomenological MIT case constant with the Dual-QCD bag term \cite{Chandola2016,Punetha2022}(either a term $B_{\text{Dual}}$ or a thermal $B(T)$)
\begin{equation}
P_{\text{Dual}}(T;V,q)= P_{BG}(T)+ P_{\text{Tsallis}}(T;V,q)+P_{\text{(Dual bag)}}.
\end{equation}
The Dual bag pressure is given by
\begin{equation}
P_{\text{(Dual bag)}}=
\begin{cases}
\displaystyle
-B_{\text{Dual}}, \ \ \ \ \ \ \ \ \text{at zero temperature},
\\[2.2ex]
\displaystyle
-B_{\text{Dual}}(T), \ \ \ \text{at finite temperature}.
\end{cases}
\end{equation}
In Dual QCD, the bag constant is related to a mass scale by:
\begin{equation}\label{bm}
    B_{\text{Dual}} =  \left(\frac{12}{\pi^2}\right)^3\ \left( c\ m \right)^4= \left(\frac{12}{\pi^2}\right)^3\ B_{\text{0}},
\end{equation}
where $m$ represents the mass of the magnetic glueballs evolving as the vector mode of the magnetically condensed vacuum \cite{Chandola2016}. The Dual QCD model also provides a thermal bag form:
\begin{equation}\label{bt}
   B(T) = B_{\text{Dual}} \left[1- 2 d \ \left(\frac{T}{m}\right)^2+d^2\left(\frac{T}{m}\right)^4\right].
\end{equation}
This corresponds to the strongly coupled regime of QCD, where the coupling is large ($\alpha_s=g^2 /4\pi \sim1$ and $m\sim 1$ \ GeV) and color confinement is maintained. The values of constants $c=\frac{1}{8} \sqrt[4]{\left(\frac{12}{\pi^2}\right)}$, and $d=\frac{(8\pi^2 + 2\pi\alpha_s)}{3}$. The Dual bag constant is used in eq.(\ref{bm}) to represent the critical condition.
\begin{equation}
a\ T_c^4 + b\ (1-q)\,V\,T_c^7 - B_{\text{Dual}}=0.
\end{equation}
The thermal Dual bag term is utilized in eq. (\ref{bt}) to represent the critical condition, allowing the critical temperature to be expressed as
\begin{equation}\label{dual3}
a^*\ T_c^4 + b(1-q)\ V\ T_c^7+ \left(B_{\text{Dual}}^*\right)^{\frac{1}{2}}\ T_c^2-B_{\text{Dual}}=0,
\end{equation}
where ($a^*=a-(d\ \frac{3}{4\pi^2})^2$) and ($B_{\text{Dual}}^*=2\frac{d}{m^2}B_{\text{Dual}}$) are new parameters in eq.(\ref{dual3}). We can observe that the value of $B^*_{\text{Dual}}$ depends on the values of both $m$ and $\alpha_s$. In the dual bag model, the bag constant $B^{\frac{1}{4}}_{\text{Dual}}$ is not introduced as a purely phenomenological parameter but is related to the vector glueball mode of the magnetically condensed QCD vacuum. This formulation connects confinement directly to the non-perturbative vacuum structure of dual QCD. For a fixed bag parameter, the model predicts a first-order QGP phase transition, whereas a temperature-dependent $B^{\frac{1}{4}}_{\text{Dual}}(T)$, induced by the thermal evolution of the glueball mass, reduces the bag pressure and softens the transition into a rapid crossover.
\section{Results}\label{sec:3}
In fig.(\ref{figb1}) shows the total pressure $P_T/T^4$ as a function of temperature and $q$ at $V=100\ fm^3$. In both subfigures, the pressure increases with increasing temperature and decreasing nonextensivity parameter, indicating a growing effect of nonextensive contribution. The finite chemical potential in fig.(b), $\mu=0.5$\ GeV, also increases the pressure, especially at lower temperatures where $\mu/T$ is larger.\\
Fig.(\ref{figb11}) displays the pressure dependence $P^{\text{res}}(T;V,q)/T^4$ on temperature and $(1-q)$ at $V=100\ fm^3$. In the subgraph (a), corresponding to $(0.001 \le (1-q) \le 0.99)$, the contour structure is dominated by the presence of a pole, indicated by the white curve, where the denominator of the analytic expression vanishes. As we approach this line, the lines (contours) become more compressed, revealing a very rapid change in the resumed pressure and indicating singular behavior. On one side of the pole, the quantity becomes strongly positive, while on the other side, its sign changes to negative, reflecting the divergence associated with the pole condition. By contrast, the subplot (b), plotted for the period $(1.001 \le (1-q) \le 1.2)$, lies entirely outside the singular region. In this case, no pole appears within the plotted range, and the recovered pressure remains limited and smooth throughout the range. The contour lines show a regular and monotonous decrease with increasing temperature, indicating that the system is continuously evolving within this range of parameters without encountering a critical limit. Looking at the two subplots together, they clearly distinguish the singular regime, where the resumed pressure is strongly and non-analytically enhanced, from the regular regime, where its behavior remains well over the entire temperature range.
\begin{figure}[h]
 \centering
 \includegraphics[width=350pt]{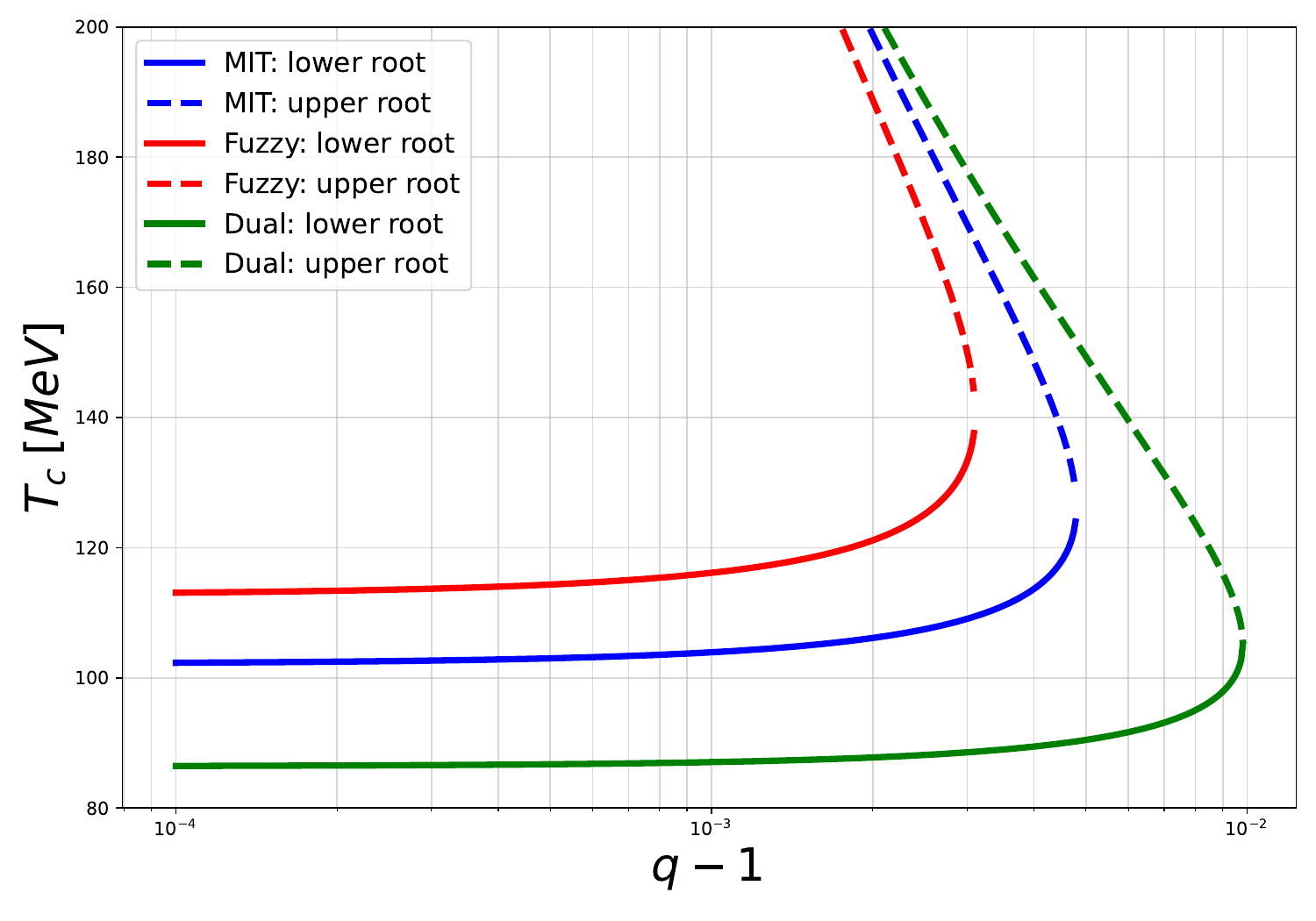}
  \caption{Critical temperature plot $T_c$ versus $(q-1)$ for MIT, Fuzzy, and Dual bag models at $V=100\ \text{fm}^3$. Solid and dashed lines indicate physical and unphysical branches, respectively.}\label{figb5}
\end{figure}
When $q=1$, the factor $(1-q)$ disappears, and consequently, the term $T_c^7$ disappears from all eqs.(\ref{MIT3}), (\ref{Fuzzytc}), and (\ref{dual3}), which represent three phenomenological models. The MIT model becomes a quartic for $T_c$, while the Fuzzy and dual models contain a quartic and a quadratic term for $T_c$. These can be solved analytically. For $q=0$, the coefficient of the $T_c^7$ term is positive, so the full nonlinear equations must be solved numerically. On the other hand, when $q>1$, one has $1-q<0$, and hence the $T_c^7$ contribution becomes negative. Using the values parameters, the numerical analysis indicates that a positive real solution for $T_c$ exists only in a narrow range above $q=1$, and no positive physical root exists outside this range (see fig.(\ref{figb5})). This behavior indicates that the sign of the $(1-q)$ term plays a crucial role in determining the existence and stability of critical temperature solutions. In this case, $T_c$ is defined as the positive real solution of a polynomial equation in temperature, with coefficients fixed by the corresponding model parameters.
\begin{table}[h!]
\centering
\caption{The values of $T^{\text{Model/w}}_c$ [MeV]) from polynomial root for $q=0,\ \text{and}\ q\geq1$ at $V=100\ \text{fm}^3$.}
\begin{tabular}{|c|c|c|c|c|}
\hline
$q$ & $T^{\text{MIT}}_c$ & $T^{\text{Fuzzy}}_c$  & $T^{\text{Dual}}_c$& $T^{\text{w}}_c$ \\
\hline
0      & 55.759  & 56.727  & 53.302 &  56 \\
1      & 102.150 & 112.789 & 86.397 & 102 \\
1.0001 & 102.312 & 113.079 & 86.459 & 102 \\
1.0005 & 102.993 & 114.328 & 87.715 & 103 \\
1.001  & 103.921 & 116.131 & 87.747 & 104 \\
1.002  & 106.122 & 121.124 & 87.760 & 107 \\
\hline
\end{tabular}
\label{tab:Tc_values}
\end{table}
For each value of $q$, the polynomial roots were computed numerically, and only positive real roots were retained, as shown in the table (\ref{tab:Tc_values}). The physical branch was identified by continuity, starting from the solution at $q=1$ and tracing the nearest positive real root as $q$ varies. This procedure is important because the coefficient of the $T^7$ term is proportional to $(1-q)$; hence, the polynomial reduces from seventh order to fourth order exactly at the Boltzmann limit $q=1$. As a result, the branch structure can change across this point, and the continuity-based formula provides a more stable determination of the physical solution than independently selecting the smallest positive root at each range point. Some works predicted that the pressure follows a $P \propto T^7$ scaling law, rather than the standard $T^4$ relationship characteristic of an ideal SB gas \cite{Miller2007,Khadkikar1993}. The study was performed over the interval ($0.99\leq q \leq1.002$), with the numerical range chosen to include exactly $q=1$. The common numerical root-tracking domain was then defined as the subset of $q$ in which the traced branch is found in all three models simultaneously.

Here, the value $q=0$ is retained as the formal criterion for the algebraic structure of the pressure equation, not as the phenomenologically favored nonextensive parameter for the QCD deconfinement transition. The corresponding root near $56$ MeV is interpreted, through comparison with hadronic models, as a nuclear liquid-gas scale rather than a QGP deconfinement scale \cite{Samanta2018}. Conversely, the conclusion that the relevant deconfinement solutions lie close to $ q=1$ is conditional on the bag parameters, finite volume, transition criterion, and the observable used to constrain $q$. Cosmological bounds and proton-proton fits should therefore be regarded as independent constraints from different systems rather than as a single universal determination of $q$ \cite{Torres1997,Sahoo2021}.

\begin{figure}[h]
 \centering
 \includegraphics[width=350pt]{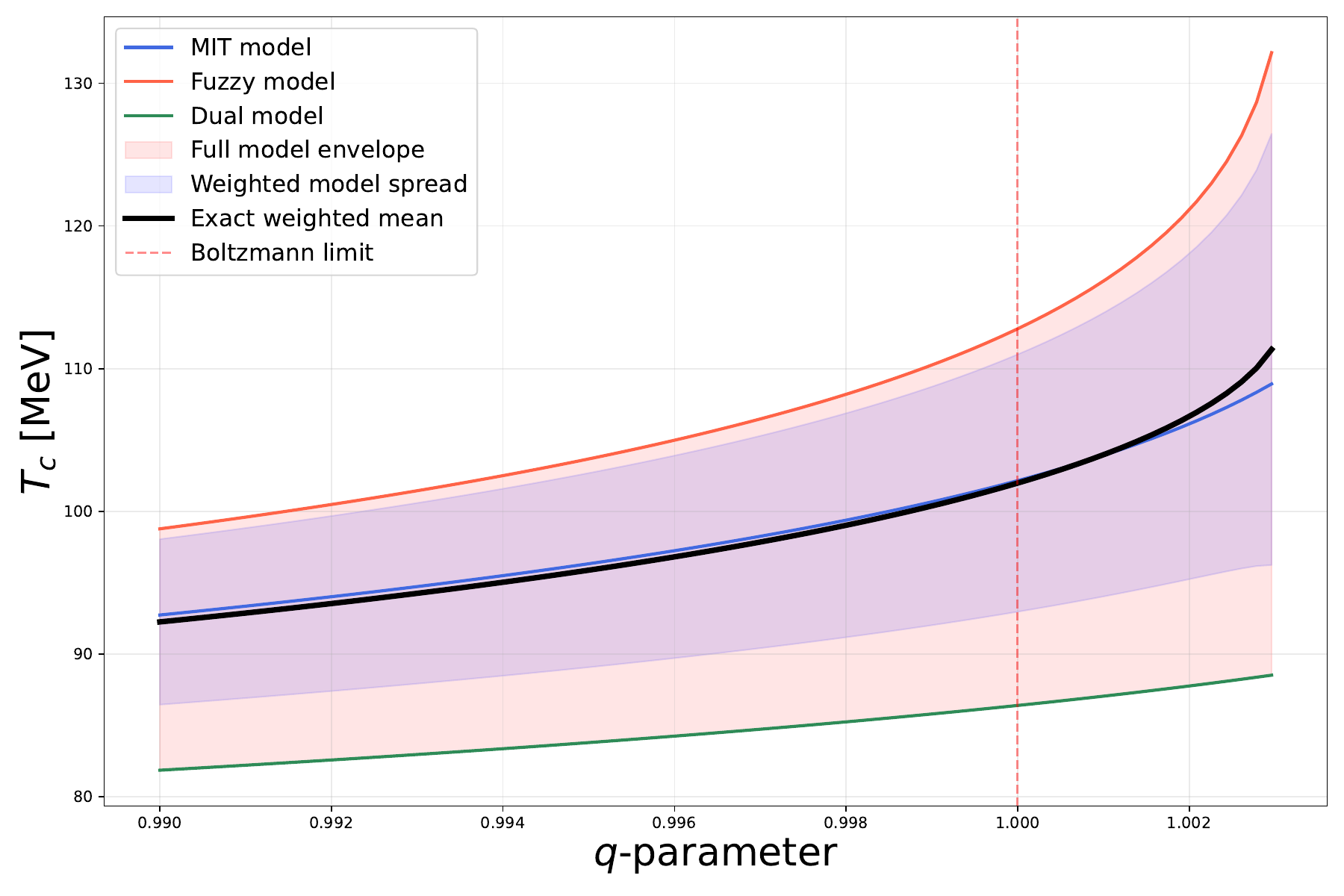}
  \caption{Weighted average and uncertainty bands for $T_c$ as a function of the non-extensive $q$-parameter, synthesized from MIT, Fuzzy, and Dual models. The dashed line marks the Boltzmann limit $q=1$.}\label{figb6}
\end{figure}

To construct a single effective prediction, the results of three models were combined using a weighted average \cite{Yang2026}. The numerical analysis in fig.(\ref{figb6}) shows that the physical branches of the three models coexist within the range ($0.99 \leq q \leq 1.01$). This range, therefore, defines the common domain in which a meaningful weighted synthesis of the three descriptions can be constructed. At the Boltzmann limit, where $q=1$, the three models yield the following critical temperatures: $T_c^{\text{MIT}}= 102.15$\ MeV, $T_c^{\text{Fuzzy}}=112.79$\ MeV, and $T_c^{\text{Dual}}= 86.4$\ MeV. Using the specified weights, the corresponding composite estimate becomes $T_c^{\text{w}}\approx 102$\ MeV, which is close to the result mentioned in the references \cite{Mhamed2019,Agasian2008}. These results indicate a clear hierarchy among the three models. The Fuzzy model predicts the highest critical temperature, the dual model gives the lowest value, and the MIT model lies between them. Since the MIT contribution dominates the weighted average, the resulting curve remains close to the MIT prediction throughout the common physical interval while still incorporating the upward shift associated with the Fuzzy model and the downward correction implied by the dual model.

To assess the sensitivity associated with different formulations for the nonperturbative pressure, we present a composite critical temperature fixed by MIT:
\begin{equation}\label{eq:Tc_weighted}
T_{\mathrm{c}}^{\mathrm{w}}(q) = \sum_{i \in \mathcal{M}} w_i T_{\mathrm{c}}^{(i)}(q),
\end{equation}
where $\mathcal{M}=\{\text{MIT}, \text{Fuzzy}, \text{Dual}\}$ denotes the set of nonperturbative pressure prescriptions, and the normalized weights ($\sum_{i \in \mathcal{M}} w_i = 1$) are fixed as follows. The weight $w_i$ depends on the chosen model and should be viewed as an estimate rather than a physical prediction. In this phenomenological model comparison scheme, the MIT model is considered as a reference description and receives the greatest weight, while the Fuzzy and Dual boundaries explain the systematic shifts resulting from alternative formulations of nonperturbative pressure. Importantly, these weights are determined independently of the resulting values of $T_{\text{c}}$ and are not derived from matching the critical temperature. Therefore, eq.(\ref{eq:Tc_weighted}) is somewhat similar to thermodynamic mixing approaches such as those in Ref. \cite{Yang2026}, where phase fractions are obtained by minimizing a common grand potential, and its coefficients do not represent phase fractions, statistical probabilities, or volume fractions. In the absence of an explicit statistical probability model or uncertainty model, $T_{c}^{w}$ is conservatively interpreted as the MIT-associated model propagation estimator, which illustrates the systematic dependence of the model on the inferred transition temperature, rather than being a thermodynamic mixture or a statistically ideal critical temperature.

A notable feature of the calculation is that all three models have one positive real solution at $q=1$, while an additional positive branch appears at $q>1$. This behavior reflects the structural change in polynomial equations resulting from multiplying the factor $(1-q)$ by the highest-order term. The appearance of a second branch directly above the Boltzmann term demonstrates that non-extensive extension modifies the solution structure even with small deviations from $q=1$. However, for $q>1$, the higher root occurs where the negative $T^7$ correction is no longer perturbative and the entropy becomes negative. We therefore interpret this root as an artifact of extrapolating the pressure ansatz beyond its range of validity, rather than as a genuine physical transition, and retain only the lower branch. The continuity of the positive branch is not uniform across the three models. Specifically, the Fuzzy model loses positive root continuity first, limiting the common period to the upper limit of ($0.99\leq q \leq 1.01$). For this reason, the weighted average should only be considered reliable within the range where the three tracked branches coexist.
\section{Conclusions}\label{sec:4}
We derived the pressure from Tsallis entropy, including the mixed term, and obtained explicit integration constants and thermodynamic quantities for MIT bag, Fuzzy bag, and Dual QCD models. The standard MIT bag model provides a clear physical basis by treating confinement as a constant vacuum pressure, it lacks the dynamic sensitivity required to describe the transition region. In contrast, the Fuzzy and Dual models introduce temperature-dependent corrections: the former is ostensibly driven by anomalies of the effect and lattice results, while the latter is theoretically driven by magnetic symmetry and the topological structure of the QCD vacuum.

The analysis of three models showed that, for $q=0$, the three models are monotonic and admit a unique positive root, while for $q=1$, the seventh-order contribution disappears and explicit analytical solutions are obtained. However, the situation is different when $q>1$, the negative $T^7$-term changes the qualitative behavior of the models and may lead to multiple positive roots or to the absence of a physical solution \cite{Vanesa2025}. When two roots occur, the higher root lies in a region where the \(T^7\) correction is no longer perturbative and the entropy becomes negative; we therefore interpret it as an artifact of extrapolating the pressure ansatz beyond its validity range, rather than as a genuine physical transition. There are studies that have indicated that the temperature has an order higher than the quartic order, $T^4$ \cite{Miller2007,Khadkikar1993}.

Studies of blackbody radiation and primordial helium constrain the cosmological value of $q$ to lie very close to unity. Blackbody-radiation analyses initially obtained $0.999964<q<1.000036$ \cite{Tsallis1995}, while primordial-helium abundances yielded the tighter bound $\lvert q-1\rvert<2.08\times10^{-5}$, corresponding to $0.9999792<q<1.0000208$ \cite{Torres1997}. These constraints apply specifically to early-universe observables and are not imposed on the present hadronic framework. Accordingly, the value $q\sim1.001$ used in proton-proton collision studies should be regarded as a system-specific near-equilibrium fit parameter rather than a cosmological value \cite{Sahoo2021}.

In the present Tsallis framework, $q$ is treated as a system-dependent phenomenological parameter: $q=1$ recovers standard BG statistics, whereas $q>1$ and $q<1$ describe correlated and anticorrelated regimes, respectively \cite{Tsallis1988,Lima2020,Sahoo2021}. Consequently, $0.99\leq q\leq1.002$ is used as a near-Boltzmann numerical interval in which the roots can be consistently traced across both regimes; it should not be interpreted as an independently measured physical bound on $q$.

This restriction to near-unity values is specific to the observables and phenomenological framework considered here, where the prescribed near-extensive deformation is introduced through a finite-volume pressure formulation and propagated through different bag-model equations. These values should therefore not be directly compared with $q$-values derived from self-consistent microscopic approaches based on self-similarity, scale invariance, or fractal hadronic spectra, in which $q$ is determined from different dynamical information \cite{Deppman2012,Deppman2014,Deppman2018,Andrade2020}. The two viewpoints are complementary: microscopic approaches investigate the possible dynamical origins of nonextensivity, whereas the present work examines how a prescribed near-extensive deformation propagates through different bag-model pressure equations.

The critical temperature $T^{\text{Model}}_c$ depends on the parameters $a$, $a^*$, $b$, $B_{\text{Model}}$ and the model components. Therefore, $T_c$ is different for each model, when $q>1$, the extracted temperatures ($102.3\leq T_c^{\text{MIT}}\leq106.1$, $\ 86.5\leq T_c^{\text{Dual}}\leq 87.8$, and $113.1\leq T_c^{\text{Fuzzy}}\leq 121.1$) are consistent with the range of QCD crossover / deconfinement scales reported in phenomenological and lattice-inspired models \cite{Chodos1974,Dumitru2012}. We received the Tsallis limit when the $q$-parameter vanished; in this case, we obtained the much lower value $T_c\sim56$ MeV, which agrees with hadron resonance gas studies, that include van der Waals interactions, where the critical point corresponds to the nuclear liquid gas transition rather than deconfinement \cite{Samanta2018,Vovchenko2015,Vovchenko2018}.

The weighted average in this analysis provides a controlled comparison of three phenomenological descriptions of the critical temperature and combines them into a single weighted estimate. In this work, the weighted average for reproducing the MIT bag model is constrained to the Boltzmann limit, $q=1$. Therefore, the resulting $T_c$ should be interpreted as the effective $T_c$ associated with the MIT model. Fuzzy and Dual contributions are included as model propagation corrections, but their weights are chosen so that their net contributions vanish at $q=1$. The result, $T_c^{w} \approx 102$\ MeV, provides a typical effective critical temperature at the Boltzmann limit. At the same time, the interval $(0.99\leq q\leq 1.002)$ defines the range within which the three models accept a common physical explanation. This establishes the $q=1$ neighborhood as the most reliable region for comparing the models and for deriving a composite estimate of the critical temperature.
\appendix
\section{}
\label{app:roots}
In this appendix, we summarize the determination of the positive roots of the critical temperature equations. Throughout this process, we assume that $T_c>0$, and that all model parameters are positive with $B^{\frac{1}{4}}\approx 145$\ MeV, $B_{Fuzzy}^{\frac{1}{2}}\approx 130\ \text{MeV}^2$, $B^{\frac{1}{4}}_0\approx 122$ MeV, $B^{\frac{1}{4}}_{\text{Dual}}\approx 141.2\ \text{MeV}$, and $\sqrt{B^{*}_{\text{Dual}}}\approx 161.8\ \text{MeV}^2$. The remaining parameters are: $a\approx 4.06$, $a^*\approx 3.627$, $b\approx 18.47$, $c\approx 0.13126$, and $d\approx 28.41334$.
\subsection{MIT bag model}
Let $q=0$ in eq.(\ref{MITc}) becomes
\begin{equation}
a\ +b\ VT_c^3-\frac{B}{T_c^4}=0.
\end{equation}
One finds
\begin{equation}
\partial_T P_{\text{MIT}}(T_c,V)=4a\ T_c^3+7b\ VT_c^6>0 \qquad (T_c>0).
\end{equation}
Hence, $P_{\text{MIT}}(T_c,V)$ is strictly increasing, with $P_{\text{MIT}}(T_c=0,V)=-B<0$ and $P_{\text{MIT}}(T_c,V)\to+\infty$ as $T_c\to\infty$. Therefore, eq.(\ref{MITc}) admits a unique positive root. When $q=1$, the $T_c^7$-term vanishes and eq.(\ref{MITc}) reduces to
\begin{equation}
aT_c^4-B=0, \Rightarrow T_c=\left(\frac{B}{a}\right)^{1/4}\approx 0.705\ B^{1/4}.
\end{equation}
To study the eq.(\ref{MITc}) when $q > 1$, we consider that $X=b\ (q-1)V>0$ becomes
\begin{equation}
a\ T_c^4-X\ T_c^7-B=0.
\end{equation}
Its derivative is
\begin{equation}
\partial_T P_{\text{MIT}}(T_c,V)=T_c^3\left(4a-7X\ T_c^3\right),
\end{equation}
so the function has a single critical point at
\begin{equation}
T_{\text{max}}=\left(\frac{4a}{7X}\right)^{1/3},
\end{equation}
Therefore, the number of positive roots is determined by the sign of $P_{\text{MIT}}(T_{\text{max}},V)$: two positive roots if $P_{\text{MIT}}(T_{\text{max}},V)>0$, one double root if $P_{\text{MIT}}(T_{\text{max}},V)=0$, and no positive root if $P_{\text{MIT}}(T_{\text{max}},V)<0$.
\subsection{Fuzzy and Dual bag models}
We now discuss the critical temperature in eqs.(\ref{Fuzzytc}) and (\ref{dual3}) for the Fuzzy and Dual models for the cases $q=0,1$. When $q=0$, the Fuzzy and Dual models give
\begin{equation}
\frac{P_{(\text{Fuzzy,Dual})}(T_c,V)}{T_c^4}=
\begin{cases}
\displaystyle
a\ +b\ VT_c^3-\frac{\sqrt{B_{\text{Fuzzy}}}}{T_c^2}-\frac{B}{T_c^4}=0,
\\[2.0ex]
\displaystyle
a^{*}+b\ VT_c^3+\frac{\sqrt{B^{*}_{\text{Dual}}}}{T_c^2}-\frac{B_{\text{Dual}}}{T_c^4}=0.
\end{cases}
\end{equation}
Its derivative is
\begin{equation}
\partial_T P_{(\text{Fuzzy,Dual})}(T_c,V)
=
\begin{cases}
\displaystyle
T_c\left(4a\ T_c^2+7b\ VT_c^5-2\sqrt{B_{\text{Fuzzy}}}\right),
\\[2.0ex]
\displaystyle
T_c \left(4a^{*}T_c^2+7bVT_c^5+2\sqrt{B^{*}_{\text{Dual}}}\right).
\end{cases}
\end{equation}
The term in parentheses is strictly increasing for $T_c>0$. Therefore, $P_{(\text{Fuzzy,Dual})}(T_c,V)$ has only one extremum. Since $P_{(\text{Fuzzy,Dual})}(0,V)=-B<0$ and $P_{(\text{Fuzzy,Dual})}(T_c,V)\to+\infty$ for large $T_c$, the equation has a unique positive root.
For $q=1$, the critical temperature in eqs.(\ref{Fuzzytc}) and (\ref{dual3}) reduce to quadratic equations in $T_c^2$, then one obtains
\begin{equation}
\frac{P_{(\text{Fuzzy,Dual})}(T_c,V)}{T_c^4}=
\begin{cases}
\displaystyle
a\ -\frac{\sqrt{B_{\text{Fuzzy}}}}{T_c^2}-\frac{B}{T_c^4}=0,
\\[2.0ex]
\displaystyle
a^{*}+\frac{\sqrt{B^{*}_{\text{Dual}}}}{T_c^2}-\frac{B_{\text{Dual}}}{T_c^4}=0.
\end{cases}
\end{equation}
Therefore, the critical temperature can be written as
\begin{equation}
T^{(\text{Fuzzy,Dual})}_c
=
\begin{cases}
\displaystyle
\left(
\frac{\sqrt{B_{\text{Fuzzy}}}+\sqrt{B_{\text{Fuzzy}}+4aB}}{2a}
\right)^{1/2},
\\[2.0ex]
\displaystyle
\left(
\frac{-\sqrt{B^{*}_{\text{Dual}}}
+\sqrt{B_{\text{Dual}}^{*}+4a^{*}B_{\text{Dual}}}}{2a^{*}}
\right)^{1/2}.
\end{cases}
\end{equation}
For $q>1$, the critical temperature in eqs.(\ref{Fuzzytc}) and (\ref{dual3}) for the Fuzzy and Dual models
can be written as
\begin{equation}
\frac{P_{(\text{Fuzzy,Dual})}(T_c,V)}{T_c^4}=
\begin{cases}
\displaystyle
a-XT_c^3-\frac{\sqrt{B_{\text{Fuzzy}}}}{T_c^2}-\frac{B}{T_c^4}=0,
\\[2.0ex]
\displaystyle
a^{*}-XT_c^3+\frac{\sqrt{B^{*}_{\text{Dual}}}}{T_c^2}-\frac{B_{\text{Dual}}}{T_c^4}=0.
\end{cases}
\end{equation}
The corresponding derivatives are
\begin{equation}
\partial_T P_{(\text{Fuzzy,Dual})}(T_c,V)
=
\begin{cases}
\displaystyle
T_c\left(4aT_c^2-7XT_c^5-2\sqrt{B_{\text{Fuzzy}}}\right),
\\[2.0ex]
\displaystyle
T_c\left(4a^{*}T_c^2-7XT_c^5+2\sqrt{B^{*}_{\text{Dual}}}\right).
\end{cases}
\end{equation}
In the both models, the negative $7^{th}$-order term dominates at sufficiently large $T_c$-values. Therefore, the pressure function can reach a local maximum. As a result, the above equation may admit two positive roots, one double positive root, or no positive root, depending on the sign of the corresponding maximum.


\begin{thebibliography}{}
\bibitem{Gross2023} F. Gross, et al., Eur. Phys. J. C \textbf{83}, 1125 (2023).
\bibitem{Gross1973} J. Gross, and F. Wilczek, Phys. Rev. Lett \textbf{30}, 1343 (1973).
\bibitem{Politzer1974} D. Politzer, Phys. Rep. \textbf{14}, 129-180 (1974).
\bibitem{Fodor2014} S. Bors\'{a}nyi, et al., Phys. Rev. Lett \textbf{113}, 052301 (2014).
\bibitem{Ding2015} T. Ding, F. Karsch, and S. Mukherjee, Int. J. Mod. Phys. E \textbf{24}, 1530007 (2015).
\bibitem{Busza2018} W. Busza, K. Rajagopal, and W. Der Schee, Ann. Rev. Nucl. Part. Sc. \textbf{68}, 339-376 (2018).
\bibitem{Schenke2020} B. Schenke, C. Shen, and D.Teaney, Phys. Rev. C \textbf{102}, 034905 (2020).
\bibitem{Arslandok2023} M. Arslandok, et al., Hot QCD white paper, arXiv:2303.17254 (2023).
\bibitem{Shao2010} M. Shao, et al., J. Phys. G \textbf{37}, 085104 (2010).
\bibitem{Marques2015} L. Marques, J. Cleymans, and A. Deppman, Phys. Rev. D \textbf{91}, 054025 (2015).
\bibitem{Kyan2022} K. Kyan, and A. Monnai, Phys. Rev. D \textbf{106}, 054004 (2022).
\bibitem{Vanesa2025} F. Vanesa, et al., Entropy \textbf{27}, 681 (2025).
\bibitem{Tsallis1988} C. Tsallis, J. Stat. Phys \textbf{52}, 479-487 (1988).
\bibitem{Lima2020} J. Lima and A. Deppman, Phys. Rev E \textbf{101}, 040102 (2020).
\bibitem{Deppman2012} A. Deppman, Physica A \textbf{391}, 6380-6385 (2012).
\bibitem{Deppman2014} A. Deppman, J. Phys. G: Nucl. Part. Phys. \textbf{41}, 055108 (2014).
\bibitem{Deppman2018} A. Deppman, et al., Entropy \textbf{20}, 633 (2018).
\bibitem{Andrade2020} E. Andrade, et al., Phys. Rev. D \textbf{101}, 054022 (2020).
\bibitem{Tirnakli2016} U. Tirnakli, and E. P. Borges, Scient. Rep. \textbf{6}, 23644 (2016).
\bibitem{Tsallis2009} C. Tsallis, Eur. Phys. J. A \textbf{40}, 257 (2009).
\bibitem{Astorga2024} M. Astorga, and G. Corral, Entropy \textbf{26}, 183 (2024).
\bibitem{Huang2023} M. Huang, and P. Zhuang, Symmetry \textbf{15}, 541 (2023).
\bibitem{Lugones2023} G. Lugones, and A. Grunfeld, Phys. Rev. D \textbf{107}, 043025 (2023).
\bibitem{Fraga2024} S. Fraga, L. Palhares, and T. Restrepo, Phys. Rev. D. \textbf{109}, 054033 (2024).
\bibitem{Chodos1974} A. Chodos, et al., Phys. Rev. D \textbf{9}, 3471 (1974).
\bibitem{DeGrand1975} T. DeGrand, et al., Phys. Rev. D \textbf{12}, 2060 (1975).
\bibitem{Philipsen2013} O. Philipsen, Prog. Part. Nucl. Phys. \textbf{70}, 55-107 (2013).
\bibitem{Karsch2005} F. Karsch, J. Phys. G: Nucl. Part. Phys \textbf{31}, S633-S640 (2005).
\bibitem{Cleymans1986} J. Cleymans, R. Gavai and E. Suhonen, Phys. Rep. \textbf{130}, 217-292 (1996).
\bibitem{Gasser1987} J. Gasser and H. Leutwyler, Phys. Lett. B \textbf{188}, 477-481 (1987).
\bibitem{Teweldeberhan2003} A. Teweldeberhan, H. Miller, and R. Tegen, Int. J. Mod. Phys. E \textbf{12}, 395 (2003).
\bibitem{Andreev2007} O. Andreev, Phys. Rev. D \textbf{76}, 087702 (2007).
\bibitem{Pisarski2007} R. Pisarski, Prog. of Theo. Phys. Suppl \textbf{168}, 276-284 (2007).
\bibitem{Chandola2009} H. Chandola, and D. Yadav, Nucl. Phys. A \textbf{829}, 151-169 (2009).
\bibitem{Punetha2022} G. Punetha, A. Bandyopadhyay, and S. Bisht, Int. J. of Mod. Phys. A \textbf{37}, 2250175 (2022).
\bibitem{Chandola2016}H. Chandola, G. Punetha, and H. Dehnen, Nucl. Phys. A \textbf{945}, 226-247 (2016).
\bibitem{Dragoa2004} A. Dragoa, A. Lavagno, and P. Quarati, Physica A \textbf{344}, 472-477 (2004).
\bibitem{Lavagno2010} A. Lavagno, D. Pigato, and P. Quarati, J. Phys. G \textbf{37}, 115102 (2010).
\bibitem{Joshi2021} S. Joshi, S. Sau, and S. Sanyal, J. High Energy Astrophys \textbf{30}, 16-23 (2021).
\bibitem{Hasmik2025} S. Hasmik, T. Sargsyan, and A. Babajanyan  , Nucl. Phys. B \textbf{1022}, 117277 (2026).
\bibitem{Swendsen2017a} R. Swendsen, Entropy \textbf{19}, 603 (2017).
\bibitem{Swendsen2017b} R. Swendsen, Physica A \textbf{467}, 67-73 (2017).
\bibitem{KAPUSTA} J. Kapusta, Finite-Temperature Field Theory Principles and Applications, Cambridge Publishing Company, (2006).

\bibitem{Mhamed2021} S. Cherif, et al., Phys. Scr. \textbf{96}, 105302 (2021) \href{https://doi.org/10.1088/1402-4896/ac0d31}.
\bibitem{Mhamed2026} M. A. A. Ahmed, and H. Zainuddin, Nucl. Phys. B \textbf{1025}, 117386 (2026) \href{https://doi.org/10.1016/j.nuclphysb.2026.117386}.
\bibitem{Mhamed2026a} M. A. A. Ahmed, and H. Zainuddin, Phys. Scr. \textbf{101} 235102 (2026) \href{https://doi.org/10.1088/1402-4896/ae7669}.
\bibitem{Mameda2023} K. Mameda, and K. Takizawa, Phys. Lett. B \textbf{847}, 138317 (2023).
\bibitem{Fraga2009} E. Fraga, L. Palhares, and C. Villavicencio, Phys. Rev. D \textbf{79}, 014021 (2009).
\bibitem{Miller2007} D. Miller, Phys. Rep. \textbf{443}, 55-96 (2007).
\bibitem{Khadkikar1993} S. Khadkikar, J. Parikh, and P. Vinodkumar, Mod. Phys. Lett A \textbf{8}, 749-755 (1993).
\bibitem{Samanta2018} S. Samanta, and B. Mohanty, Phys. Rev. C \textbf{97}, 015201 (2018).
\bibitem{Torres1997} F. Torres, H. Vucetich, and A. Plastino, Phys. Rev. Lett \textbf{79}, 1588 (1997).
\bibitem{Sahoo2021} D. Sahu, and R. Sahoo, Physics \textbf{3}, 207-219 (2021).
\bibitem{Yang2026} Y. Yang, et al., Phys. Rev. D \textbf{113}, 114018(2026).
\bibitem{Mhamed2019} M. A. A. Ahmed, et al., Int. J. Mod. Phys. A \textbf{34}, 1950051 (2019)\href{https://doi.org/10.1142/S0217751X19500519}.
\bibitem{Agasian2008} O. Agasian, and S. Fedorov, Phys. Lett B \textbf{663}, 445-449 (2008).
\bibitem{Tsallis1995} C. Tsallis, F. Barreto, and E. Loh, Phys. Rev. E \textbf{52}, 1447 (1995).
\bibitem{Dumitru2012} A. Dumitru, et al., Phys. Rev. D \textbf{86}, 105017 (2012).
\bibitem{Vovchenko2015} V. Vovchenko, D. Anchishkin, and M. Gorenstein, Phys. Rev. C \textbf{91}, 024905 (2015).
\bibitem{Vovchenko2018} V. Vovchenko, M. Gorenstein, and H. Stoecker, Phys. Rev. lett \textbf{118}, 182301 (2017).
\end{thebibliography}
\end{document}